\documentclass[useAMS,usenatbib]{mn2e}
\usepackage{epsfig}
\usepackage{amsmath}
\usepackage{amssymb}

\newcommand{\be}{\begin{equation}}
\newcommand{\beq}{\begin{equation}}
\newcommand{\ba}{\begin{eqnarray}}
\newcommand{\ee}{\end{equation}}
\newcommand{\eeq}{\end{equation}}
\newcommand{\ea}{\end{eqnarray}}

\newcommand{\Lylim}{Lyman limit}

\newcommand{\omegab}{\Omega_{\mathrm{B}}}

\newcommand{\scs}{\scriptstyle}

\def\lsim{~\rlap{$<$}{\lower 1.0ex\hbox{$\sim$}}}
\def\gsim{~\rlap{$>$}{\lower 1.0ex\hbox{$\sim$}}}

\title[Ionizing Photon Escape Fraction]{Determining the escape fraction of ionizing photons during reionization with the GRB derived star-formation rate}

\author[Wyithe et al.]{J.S.B. Wyithe$^1$, A.M. Hopkins$^2$, M.D. Kistler$^{3,4}$, H. Y\"{u}ksel$^5$, J.F. Beacom$^{3,4,6}$\\
$^1$School of Physics, University of Melbourne, Parkville, Victoria, Australia\\
$^2$Anglo-Australian Observatory, P.O. Box 296, Epping, NSW 1710, Australia\\
$^3$Center for Cosmology and Astro-Particle Physics, The Ohio State University,
191 W. Woodruff Ave., Columbus, OH 43210\\
$^4$Dept. of Physics, The Ohio State University, 191 W. Woodruff Ave.,
Columbus, OH 43210\\
$^5$Bartol Research Institute and Department of Physics and Astronomy,
University of Delaware, Newark, Delaware 19716\\
$^6$Dept. of Astronomy, The Ohio State University, 140 W. 18th Ave.,
Columbus, OH 43210\\
email: swyithe@unimelb.edu.au}

\date{Draft Version}
\pagerange{\pageref{firstpage}--\pageref{lastpage}}
\pubyear{2006}

\begin{document}
\label{firstpage}
\maketitle
\begin{abstract}

  The fraction of ionizing photons that escape their host galaxies and
  so are able to ionize hydrogen in the inter-galactic medium (IGM) is
  a critical parameter in analyses of the reionization era and early
  galaxy formation. Studies of the reionization history normally
  suffer from a degeneracy between the unknown values for the efficiency with which high
  redshift galaxies turn mass into stars and the escape
  fraction of ionizing photons. Recent gamma-ray
  burst (GRB) measurements of the star formation rate density during reionization provide the first opportunity to break this degeneracy. We
  confront a semi-analytic model for reionization with the GRB-derived
  star formation rate, as well as observations of the Ly$\alpha$ forest and the CMB. Assuming that UV
  photons produced in star-forming galaxies dominate the reionization
  process, we show that the escape fraction of ionizing photons from
  high redshift galaxies is $\sim5\%$ [$\log{f_{\rm
      esc}}=-1.35\pm0.15$ ($68\%$)] for our fiducial model. This value is
  reasonably stable against uncertainties in the modeling, including
  the implementation of radiative feedback, the possibility of an
  evolving escape fraction, and the unknown shape of the IMF, which in
  sum contribute $\sim0.2$ dex of additional systematic uncertainty on
  the value of escape fraction.

\end{abstract}
\noindent 

\section{INTRODUCTION}
\label{introduction_section}

Star-bursting galaxies and quasars have been the leading candidates for the sources of the UV radiation required to reionize the hydrogen
gas in the inter-galactic medium (IGM) \citep[e.g.,][]{Barkana2001}.  The
quasar population is observed to decline quickly at $z\gtrsim 2.5$ and
so it is believed that it was galaxies that contributed the bulk of
UV photons that drove reionization
\citep{Madau1999,Fan2002,Srbinovsky2007,Bolton2007}.  However, it has been difficult to make a positive case for galaxies as opposed to favoring them by disfavoring the case for quasars. The contribution
of galaxies to the UV radiation field is dependent on the star
formation rate and initial stellar mass function, but is also limited by the
fraction of ionizing photons that escape their host galaxies.  If the
escape fraction is small, then star formation had to be very efficient
at high redshift in order to reionize the Universe. The escape
fraction is therefore a critical parameter in studies of  the connection between high redshift galaxy formation and reionization.

Attempts to determine the escape fraction have been dominated by
direct observations of relatively low redshift galaxies, and by
numerical simulation.  Direct measurements of the escape fraction
($f_{\rm esc}$) are complicated by the fact that the intrinsic number
of ionizing photons produced by a specific galaxy is unknown.  A
commonly adopted strategy to circumvent this limitation is to compare
the flux observed at the Lyman limit to the observed flux at a
frequency where the intrinsic emissivity can be inferred
\citep{Leitherer1995, Heckman2001, Deharveng2001,
  Steidel2001,Siana2007}.  The escape fraction at the {\Lylim} can be
then be derived using a model description of the star formation
history \citep{Leitherer1995, Steidel2001, Siana2007}. However the
results are currently uncertain.  At redshifts $z\sim1-3$,
observations have suggested a broader range of values for $f_{\rm
  esc}$, from a few percent to $\sim 20\%$ \citep{Steidel2001,
  Fernandez2003, Shapley2006, Siana2007}. \citet{Inoue2006} have
examined the evolution of the escape fraction in the redshift range
$z=0-6$ using both direct observations of the escape fraction and
values that they derive from measurements of the ionizing background.
They find that the escape fraction evolves from $f_{\rm esc} \sim
1-10\%$, increasing towards high redshift. More recently
\citet{Fynbo2009} have used spectroscopic observations of individual
GRB afterglows at $z>2$ to place a 95\% confidence upper limit of
7.5\% on the escape fraction for ionizing photons on the sightlines of
these GRBs.

Theoretical modeling has concentrated on the absorption of ionizing
photons as they propagate through the inter-stellar medium towards the
IGM.  Here also the results are not yet conclusive. For example, \citet{Wood2000}
found $f_{\rm esc}\lesssim 1\%$ at $z\sim10$ suggesting that the escape
fraction decreases towards high redshift due to the increased density
of galactic disks. More recently, \citet{Razoumov2006} used galaxy
formation simulations incorporating high-resolution 3-D radiative
transfer to show that the escape fraction evolves from
$f_{\rm esc}\sim1-2\%$ at $z=2.39$ to $f_{\rm esc}\sim6-10\%$ at $z=3.6$.  In
agreement with \citet{Fujita2003}, \citet{Razoumov2006} find that
increased supernova feedback at higher redshift expels gas from the
vicinity of star-bursting regions, creating tunnels in the galaxy
through which ionizing photons can escape into the IGM.  They also
find that star formation may occur at slightly lower densities at
higher redshift, which further contributes to the increased $f_{\rm esc}$
towards higher redshift.

Recently, detailed numerical simulations \citep{Gnedin2007a} have been
presented that predict a value for $f_{\rm esc}$ between $1$ and $3\%$, for
halos of mass $M\gtrsim5\times10^{\scs 10}M_{\odot}$, over the
redshift range $3<z<9$.  This very low efficiency of reionization
would have profound effects on the reionization history.  In addition
to a small escape fraction in massive galaxies, \citet{Gnedin2007a}
further predict that halos with $M\lesssim 5\times10^{\scs
10}M_{\odot}$ have an escape fraction that is negligibly small. We
point the interested reader to \citet{Srbinovsky2008} and to the
review by \citet{Ciardi2005} for a more detailed discussion of prior
work on estimating the escape fraction.

A less direct avenue to determining the escape fraction is through
modeling of the ionizing background at the end of reionization. Models
of the reionization of hydrogen and the subsequent Ly$\alpha$ forest
are subject to a degeneracy between the star formation efficiency of
gas accreted into galaxies and the fraction of ionizing photons that
escape the dense gaseous environment of the galaxy to ionize the
IGM. In order to break this degeneracy, the true cosmic star formation rate must
be known. For example, the escape fraction has been estimated at
high redshifts by combining the ionization rate in the IGM with an
estimate of the star formation rate obtained from measurements of the
galaxy luminosity function \citep[e.g.,][]{Bolton2007,Srbinovsky2008,Faucher2008}. Assuming that the mean-free-path is accurately
estimated, the escape fraction follows from comparison of the total
number of photons produced with the number that are ionizing the
IGM. The primary drawback of this technique is that the observed
luminosity functions are subject to uncertain corrections for
extinction in galaxies, as well as a flux limit. The latter fact means that an
extrapolation to lower galaxy luminosities must be performed in order to
obtain a total star formation rate.

GRBs are now beginning to probe the star formation rate density in the
era during which the Universe is thought to have become reionized
\citep{Salvaterra2009,Tanvir2009}. In this paper we employ the recent
determination of the star formation rate density \citep{Kistler2009} out to $z\sim8.5$ based
on a compilation of high redshift GRBs. As described below, this measurement of the cosmic star formation rate with GRBs is not subject to the usual large corrections required for extinction and unseen faint galaxies. By
modeling the reionization of hydrogen we determine the star formation
efficiency and escape fraction that is required to simultaneously
reproduce the optical depth to electron scattering of CMB photons, the
star formation rate density during the reionization era, and the
ionizing background at the end of reionization. Inclusion of the star
formation rate density as a constraint allows the star formation
efficiency and escape fraction to be separately determined.

Our analysis is similar to that of
\citet{Choudhury2005,Choudhury2006}, who modeled the reionization
history and compared model predictions to a range of observables,
including the star formation rate, the ionization rate from the
Ly$\alpha$ forest and the optical depth to electron scattering of
CMB photons, as well as the evolution of Lyman limit systems and the temperature of the IGM among others. However, here we limit our attention to the
first three of these observables that most directly relate to
determination of the escape fraction, and which can be most reliably
predicted by an analytic model. Most importantly, we make use of the recent
determination of star formation rate density from GRBs, which for the first
time provide a measurement of the total star formation deep into the
reionization era. We also quantify the statistical and systematic
uncertainties on the model parameters.

The paper is organized as follows. In \S~\ref{GRB} we discuss the
determination of star formation rate density from high redshift GRBs, and
review a simple estimate of the ionizing photon budget for
reionization. We then describe other observational constraints
(\S~\ref{others}), as well as our semi-analytic model for reionization
(\S~\ref{models}).  The resulting limits on the possible values of
model parameters presented in \S~\ref{constraints}. We conclude with a
summary of our results for the escape fraction in
\S~\ref{conclusion}. In our numerical examples, we adopt the standard
set of cosmological parameters \citep{Komatsu2009}, with values of
$\Omega_{\rm b}=0.04$, $\Omega_{\rm m}=0.24$ and $\Omega_\Lambda=0.76$ for
the matter, baryon, and dark energy fractional density respectively,
$h=0.73$, for the dimensionless Hubble constant, and $\sigma_8=0.81$
for the variance of the linear density field within regions of radius
$8h^{-1}$Mpc.

\section{A GRB determination of the star formation rate density at high redshift}
\label{GRB}

\citet{Kistler2009} \citep[following][]{Yuksel2008} have used the
counts of GRBs at high redshift to infer the star formation rate density
out to $z\sim8.5$. Inspection of their Figure~4 (see also
Figure~\ref{fig1} of this paper) suggests that the star-formation rate density
is as large as or could even be higher at $z\sim8$ than at $z\sim4-6$,
implying that the ionizing photon emissivity is not falling as
observations move into the reionization era. This inference requires
an assumption for the initial mass function. However since both
ionizing photons and GRBs are produced by massive stars, estimates of
the ionizing photon emissivity from the GRB rate should be fairly
robust against uncertainties in the initial mass function (this point
is taken up quantitatively in \S~\ref{IMF}).  \citet{Kistler2009}
argued that we have observed enough star formation via high redshift
GRBs at $z\ga6$ to reionize the universe. This conclusion was reached
via a simple estimate of the number of ionizing photons produced prior
to $z\sim6$ given the observed star formation rate density. For a Salpeter initial mass
function, 4648 ionizing photons are produced per baryon incorporated
into stars (\S~\ref{IMF}). Taking this value, together with a constant
star formation rate density for a time interval $\Delta t$ the number of
photons produced by stars per baryon of IGM is
\begin{equation} 
\label{eq1}
\mathcal{N}_\gamma \sim 4 \left(\frac{f_{\rm esc}}{0.1}\right) \left(\frac{\dot{\rho}_\star}{0.1M_\odot\mbox{Mpc}^{-3}}\right) \left(\frac{\Delta t}{400\mbox{Myr}}\right)
\end{equation} 
\citep{Kistler2009}. The inferred star formation rate density at $z>6$
implies $\mathcal{N}_\gamma\sim4^{+4}_{-2}(f_{\rm esc}/0.1)$. This is
consistent with the modeling of Wyithe \& Cen~(2007) who found
$\mathcal{N}_\gamma\sim4$ for reionization concluding at $z=6$ under a
range of assumptions for the redshift range and efficiency of
Population-III star formation, provided that the escape fraction of
ionizing radiation is of order 10\%.  Thus, calculations based on the
GRB-derived star formation rate density imply that there was
sufficient star formation to complete reionization even with a
relatively small value of $f_{\rm esc}$.

In this paper we extend the simple estimates presented in
\citet{Kistler2009}. Specifically, we constrain the possible values of
escape fraction by modeling the reionization history and comparing
with the observed star formation rate density as traced by high redshift GRBs,
the ionization rate as traced by the Ly-$\alpha$ forest at $z<6$, and
the optical depth to Thomson scattering of CMB photons. The
GRB-derived star formation rate density values that were presented in
\citet{Kistler2009} are special in three ways with respect to breaking
the degeneracy between escape fraction and star formation efficiency
during the reionization era.  

Firstly, the star formation rate density is
based on number counts of GRBs which yields the absolute death/birth
rate of massive stars. These number counts are independent of the
escape fraction that can influence other star formation rate
indicators based on UV emission. The utility of counting
events marking the death of individual stars in order to get the star
formation rate density has also been argued by \cite{Lien2009} in the context of
supernovae at lower redshift. 

Secondly, GRBs are thought
to sample the full luminosity range of galaxies, so that no correction
for luminosity function is required.  In conventional studies of the
star formation rate, one attempts to correct for both problems, but
there are uncertainties \citep[e.g.,][]{Hopkins2006}. 
The GRB results from \cite{Kistler2009} confirm the two large
corrections (for obscuration and luminosity function) applied to the raw results of
\citet{Bouwens2008}. It could therefore be argued that this agreement
implies the combined corrections are approximately correct, suggesting the
absence of any deep misunderstandings about the properties of high
redshift galaxies. Conversely, if the luminosity function correction
applied to the \citet{Bouwens2008} data by \citet{Kistler2009} is appropriate, then most of the total
star formation during the reionization era was within small
galaxies. With regard to this point it is important to note that in more
local samples, most GRB hosts are found to be low mass galaxies \citep{Fruchter2006,Stanek2006}. This implies that GRBs
are an efficient probe of low mass galaxy formation, and is in
contrast to studies of the UV luminosity function at $z>6$ that currently probe
only the most massive galaxies. Since low mass galaxies seem to be
dominant during the reionization era we therefore argue that GRBs
provide the most important constraints on the star formation rate density
during reionization, and utilize the star formation values from \citet{Kistler2009}
in our parameter constraints.

The third fundamental advantage of using GRBs to determine the star
formation rate density is that the data now goes deep into the reionization
era. This makes them a rather direct probe of the star formation
responsible for reionization. Importantly, we note that even  though the
numbers of GRBs used in the estimate are low,
strong statements can still be made about the star formation rate density
\citep{Kistler2009,Yuksel2008}. Of course the addition of more high
redshift GRBs in the future will aid our understanding of the
reionization era by lowering the statistical errors on estimation of
the star formation rate density.

\subsection{Summary of star formation rate density from high-z GRBs}

\begin{figure*}
\includegraphics[width=15cm]{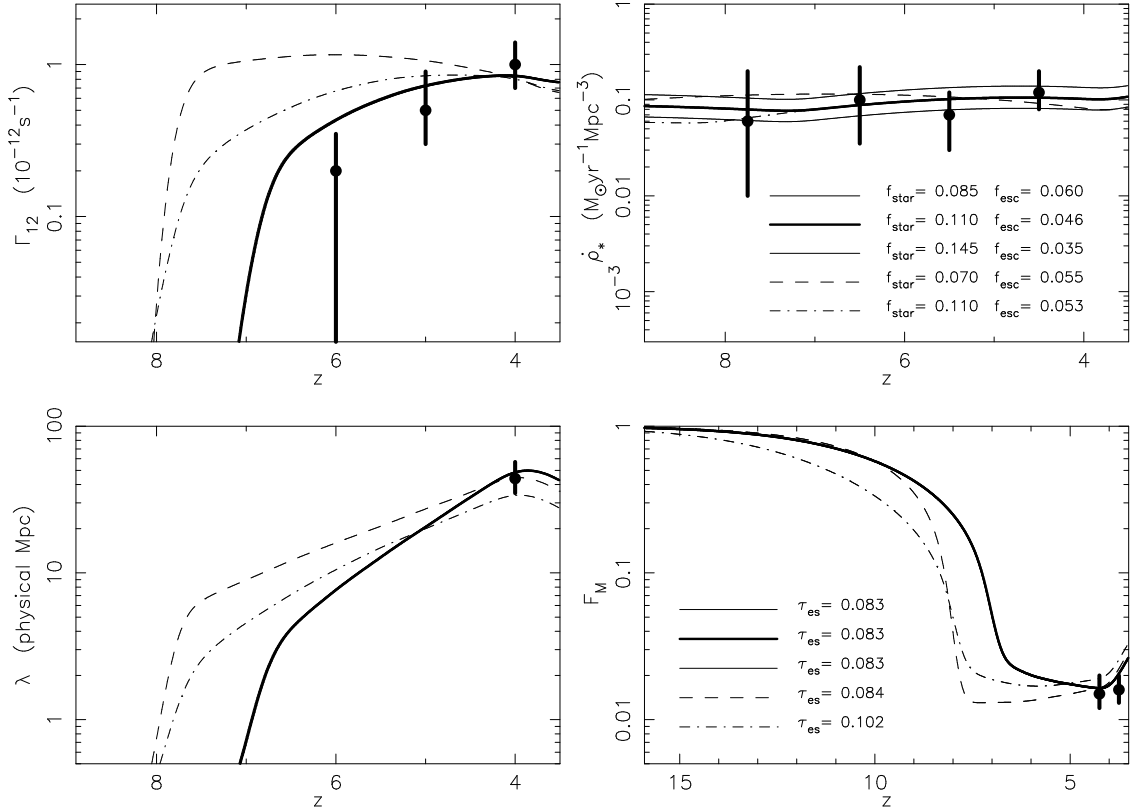} 
\caption{Models for the reionization of the IGM and the subsequent
post-overlap evolution of the ionizing radiation field. 
In each panel 5 cases are shown, with line-styles corresponding to the cases in Figures~\ref{fig2}-\ref{fig3}. The three solid lines shown correspond to the best fit (thick line) and 1-sigma deviations for the fiducial case with $T_{\rm min}=10^4$K and $T_{\rm ion}=10^5$K (note that these 3 cases for the fiducial model overlap in some of the panels). The dashed line shows a case with $T_{\rm min}=T_{\rm ion}=10^4$K. The dot-dashed line shows a case with $T_{\rm min}=10^4$K and $T_{\rm ion}=10^5$K, but $f_{\rm esc}\propto(1+z)$. The models are labeled by the values of $f_\star$, $f_{\rm esc}$ and $\tau_{\rm es}$. 
A value of $\Delta_{\rm c}=10$ was chosen for the critical overdensity prior
to the overlap epoch. The histories shown correspond to best fit models based on constraints from $\Gamma_{12}$ and $\dot{\rho}_\star$ at $z>4$, and also $\tau_{\rm es}$. {\em Upper Left Panel:} The ionization rate as a function of redshift. The observational points are from \citet{Bolton2007}. {\em Upper Right Panel:} The star formation rate density as a function of redshift. The data points are from \citet{Kistler2009}.  {\em Lower Left Panel:} The mean-free-path for ionizing photons. The data point is based on \citet{Storrie-Lombardi1994}. {\em Lower Right Panel:} The volume and mass averaged fractions of neutral gas in the universe. The observational
points for the mass-fractions are from the damped Ly$\alpha$ measurements of \citet{Prochaska2005}. 
}
\label{fig1}
\end{figure*}

As discussed above, \citet{Kistler2009} have
determined the star formation rate density deep into the reionization era
based on the occurrence of GRBs at $z>6$. We use as constraints the determination of star formation
rate density based on GRBs between $4\la z\la8.5$ from \citet{Kistler2009}. In units of solar masses per year per cubic Mpc, the values used are
$\dot{\rho}_{\rm \star,obs}=0.06^{+0.14}_{-0.05}$ at $z=7.75\pm0.75$,
$\dot{\rho}_{\rm \star,obs}=0.1^{+0.17}_{-0.07}$ at $z=6.5\pm0.5$,
$\dot{\rho}_{\rm \star,obs}=0.07^{+0.05}_{-0.04}$ at $z=5.5\pm0.5$, and
$\dot{\rho}_{\rm \star,obs}=0.13^{+0.07}_{-0.05}$ at $z=4.5\pm0.5$. When combined with
the fact that CMB constraints \citep{Komatsu2009} place most of the reionization at $z\sim10$
this implies that we now have multiple constraints probing epochs during the
reionization era. These data are shown in the upper right panel of Figure~\ref{fig1}.

\section{Additional Observational Constraints}
\label{others}

In this section we summarise two additional pieces of data that constrain the
possible scenarios in addition to the GRB derived star formation rate density. 

\subsection{Thomson scattering optical depth for CMB photons}

The measured optical depth to Thomson scattering of CMB
photons measures the column density of ionized hydrogen between the
observer and the surface of last scattering. The latest observed value
of $\tau_{\rm es}=0.084\pm0.016$ implies that reionization did not
occur significantly beyond $z\sim10$ \citep{Komatsu2009}. When computing $\tau_{\rm es}$ from the reionization history we assume that the filling factor of singly ionized helium equals the ionized fraction of hydrogen, and that helium becomes doubly ionized at $z=3$ \citep{Wyithe2003}.

\subsection{Ionization rate from the Ly$\alpha$ forest}

Observations of the Gunn-Peterson optical depth in the high redshift
Ly-$\alpha$ forest imply a very highly ionized IGM at $z\la6$,
suggesting that reionization was complete by that time. The intensity of
the ionizing background implied by the effective Ly$\alpha$ optical
depth is sensitive to the details of the distribution of gas densities
and temperatures, but can be reliably modeled via numerical simulation
\citep[e.g.,][]{Bolton2007}. We take values for the ionization rate at
$4\la z\la6$ from the simulations of \citet{Bolton2007}, based on
the observations of \citet{Fan2006a}. In units of $10^{-12}$s$^{-1}$ the values used are $\Gamma_{\rm obs}=0.2^{+0.15}_{-0.2}$ at $z=6$, $\Gamma_{\rm obs}=0.5^{+0.4}_{-0.2}$ at $z=5$, $\Gamma_{\rm obs}=1^{+0.4}_{-0.3}$ at $z=4$ respectively. These data are shown in the upper left panel of Figure~\ref{fig1}.

\section{Semi-Analytic Model for Reionization}
\label{models}

In this section we summarise the semi-analytic model used to calculate the reionization history of the IGM. The basis of this model is the differential between ionization and recombination rates for hydrogen in an inhomogeneous IGM.
\citet{Miralda2000} presented a model that allows the
calculation of an effective recombination rate in an inhomogeneous
universe by assuming a maximum overdensity ($\Delta_{\rm c}$)
penetrated by ionizing photons within HII regions. The model assumes
that reionization progresses rapidly through islands of lower density
prior to the overlap of individual cosmological ionized
regions. Following the overlap epoch, the remaining regions of high
density are gradually ionized. It is therefore hypothesized that at
any time, regions with gas below some critical overdensity
$\Delta_{\rm i}\equiv {\rho_{i}}/{\langle\rho\rangle}$ are highly
ionized while regions of higher density are not. 
The fraction of mass in regions with overdensity below $\Delta_{\rm i}$, is found from the integral
\begin{equation}
F_{\rm M}(\Delta_{\rm i})=\int_{0}^{\Delta_{\rm i}}d\Delta P_{\rm
V}(\Delta)\Delta,
\end{equation}
where $P_{\rm V}(\Delta)$ is the volume weighted probability
distribution for $\Delta$.  \citet{Miralda2000} quote a
fitting function that provides
 a good fit to the volume weighted probability distribution for the
baryon density in cosmological hydrodynamical simulations.   
In what follows, we
draw primarily from the prescription of Miralda-Escude et al.~(2000) and refer the reader to the
original paper for a detailed discussion of its motivations and
assumptions.  \citet{Wyithe2003} employed this prescription within
a semi-analytic model of reionization. This model was extended by
\citet{Srbinovsky2007} and by \citet{Wyithe2008}. We refer the reader to those papers for a full
description.

The quantity $Q_{\rm i}$ is defined to be the
volume filling factor within which all matter at densities below
$\Delta_{\rm i}$ has been ionized. The reionization history is quantified by the evolution of $Q_{\rm i}$ that evolves according to the rate equation
\begin{eqnarray}
\label{preoverlap}
\nonumber \frac{dQ_{\rm i}}{dz} &=& \frac{1}{n_0 F_{\rm M}(\Delta_{\rm
i})}\frac{dn_{\gamma}}{dz}\\ \nonumber &-&\left[\alpha_{\rm
B}(1+z)^3R(\Delta_{\rm i})n_0\frac{dt}{dz}+\frac{dF_{\rm M}(\Delta_{\rm c})}{dz}\right]\frac{Q_{\rm
i}}{F_{\rm M}(\Delta_{\rm i})},
\end{eqnarray} 
where $\alpha_{\rm B}$ is the case B recombination coefficient, $n_0$
is the comoving density of hydrogen in the mean IGM, and
$R(\Delta_{\rm i})$ is the effective clumping factor of the IGM. The
evolution is driven by the rate of emission of ionizing photons per
co-moving volume $dn_\gamma/dz$. Within this formalism, the epoch of
overlap is precisely defined as the time when $Q_{\rm i}$ reaches
unity. Prior to the overlap epoch we must solve for both $Q_{\rm i}$
and $F_{\rm M}$ (or equivalently $\Delta_{\rm i}$).  The relative
growth of these depends on the luminosity function and spatial
distribution of the sources. In this regime we assume $\Delta_{\rm i}$
to be constant with redshift before the overlap epoch and compute
results for models with values of $\Delta_{\rm i}\equiv\Delta_{\rm
  c}=10$. 

Following overlap $Q_{\rm i}=1$ and we describe the
post-overlap evolution of the IGM by computing the evolution of the
ionized mass fraction according to the equation
\begin{equation}
\label{postoverlap}
\frac{dF_{\rm M}(\Delta_{\rm i})}{dz} =
\frac{1}{n_0}\frac{dn_\gamma}{dz}-\alpha_{\rm
B}\frac{R(\Delta_{\rm i})}{a^3}n_0\frac{dt}{dz}.
\end{equation}
Note that in this post overlap regime the value of $\Delta_{\rm i}$ is the dependent variable describing the ionization state of the IGM and is solved for as a function of redshift.

\begin{figure*}
\includegraphics[width=15cm]{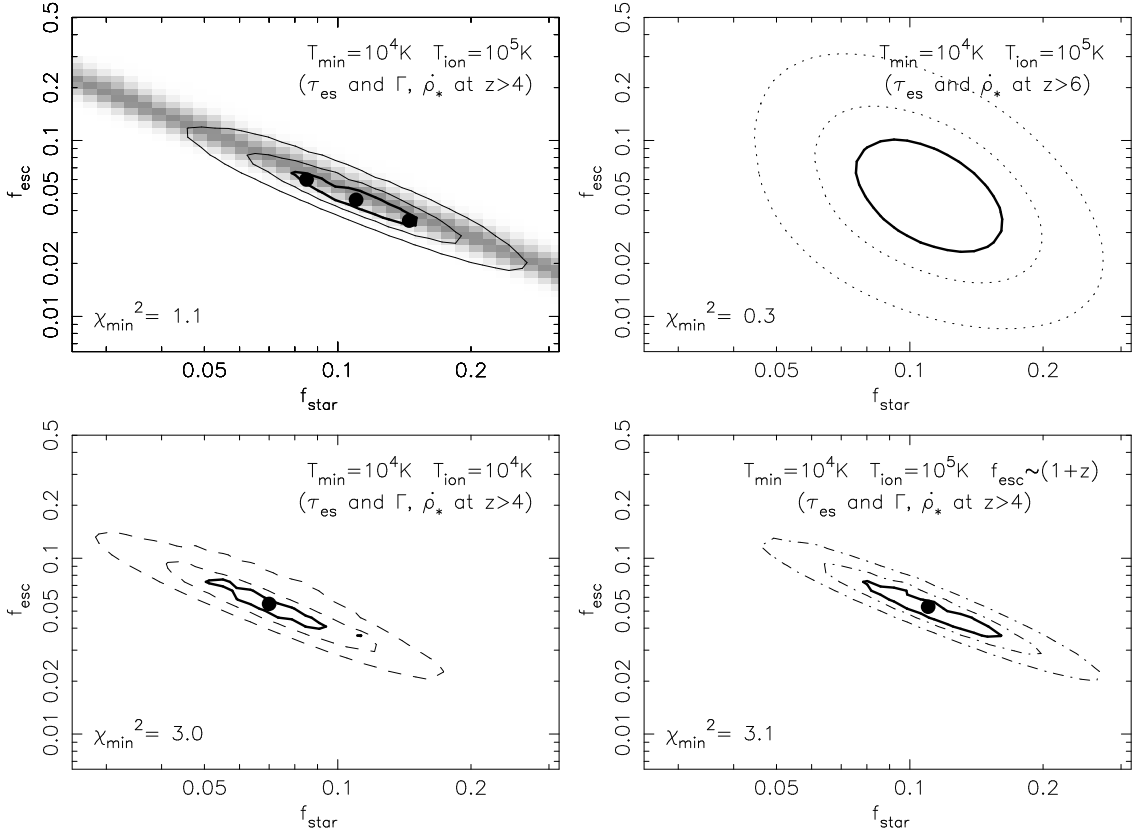} 
\caption{Joint constraints on the parameters $f_\star$ and $f_{\rm esc}$ based on observations of the star formation rate density $\dot{\rho}_\star$, the ionization rate $\Gamma_{12}$ and the optical depth to Thompson scattering of CMB photons $\tau_{\rm es}$. {\em Upper Left:} The fiducial model with $T_{\rm min}=10^4$K and $T_{\rm ion}=10^5$K. Constraints on $\Gamma_{12}$ and $\dot{\rho}_\star$ are used at $z>4$. Also shown for comparison are the constraints on this model if measurements of $\dot{\rho}_\star$ are ignored, which reveals how the GRB star formation data break the degeneracy (gray shading). {\em Upper Right:} The fiducial model with $T_{\rm min}=10^4$K and $T_{\rm ion}=10^5$K. Constraints on $\dot{\rho}_\star$ are used at $z>6$. {\em Lower Left:} A model with $T_{\rm min}=T_{\rm ion}=10^4$K. Constraints on $\Gamma_{12}$ and $\dot{\rho}_\star$ are used at $z>4$.  {\em Lower Right:} A model with $T_{\rm min}=10^4$K and $T_{\rm ion}=10^5$K, and $f_{\rm esc}\propto(1+z)$. Constraints on $\Gamma_{12}$ and $\dot{\rho}_\star$ are used at $z>4$. In each case three contours are shown corresponding to differences in $\chi^2$ relative to the best fit model of $\Delta \chi^2=\chi^2-\chi_{\rm min}^2=1$, 2.71 and 6.63. Projections of these contours onto the axes provide the 68.3\%, 90\% and 99\% confidence intervals on individual parameter values. The large dots show parameters of model histories presented in Figure~\ref{fig1}.}	
\label{fig2}
\end{figure*}

The emission rate of ionizing photons per co-moving volume that is required to compute the reionization history can be written
\begin{equation}
\label{star_ionize}
\frac{dn_{\rm \gamma}}{dz} = N_{\rm \gamma}f_{\rm
esc}\frac{\dot{\rho}_\star}{m_{\rm p}}\frac{dt}{dz},
\end{equation}
where $N_{\rm \gamma}$ is the number of ionizing photons produced per baryon incorporated into stars, and $\dot{\rho}_\star$ is the star formation rate per unit volume.
As described in the introduction, only a fraction of ionizing photons
produced by stars enter the IGM. Therefore an additional factor of
$f_{\mathrm{esc}}$ (the escape fraction) must be included when
computing the ionizing emissivity of galaxies. There are expected to be large fluctuations in escape fraction with time and with viewing angle for individual galaxies. The escape fraction could also depend on galaxy mass \citep{Gnedin2007a}. However what is important for studies of the overall photon budget during reionization is the total cosmic ionizing flux, rather than the ionizing flux near a particular galaxy (although of course the latter could affect statistics of Ly$\alpha$ absorption or 21cm emission from the IGM). In our model we therefore assume a single value for $f_{\rm esc}$, which should be interpreted as a stellar mass averaged value over star forming galaxies. 

The star formation
rate per unit volume is computed based on the collapsed fraction
obtained from the \citet{Press1974} model in halos above the
minimum halo mass for star formation, together with an assumed star
formation efficiency ($f_\star$)
\begin{equation}
\label{SFR_equation}
\dot{\rho}_\star=f_\star \rho_{\rm c}\omegab\left((1-Q_{\rm i})\frac{dF_{\rm col}(M_{\rm min})}{dt} + Q_{\rm i}\frac{dF_{\rm col}(M_{\rm ion})}{dt}\right),
\end{equation}
\noindent
where $\rho_c$ is the critical density and the collapsed
fraction includes separate components from ionized and neutral regions
of IGM. In a cold neutral IGM beyond the redshift of reionization, the
collapsed fraction should be computed for halos of sufficient mass to
initiate star formation. The critical virial temperature is set by the
temperature ($T_{{\mathrm{min}}}\sim 10^4$ K) above which efficient
atomic hydrogen cooling promotes star formation. Following the
reionization of a region, the Jeans mass in the heated IGM limits
accretion to halos above $T_{{\mathrm{ion}}}\sim10^5$ K
\citep{Efstathiou1992, Thoul1996, Dijkstra2004b}.
Thus, once $N_{\gamma}$, $T_{{\mathrm{min}}}$ and $T_{{\mathrm{ion}}}$ are specified, the reionization model has 2 free parameters $f_\star$ and $f_{\rm esc}$.

In order to estimate the ionizing background following the end of
reionization, our approach is to compute a reionization history given a
particular value of $\Delta_{\rm c}$, combined with assumed values for
the efficiency of star-formation and the fraction of ionizing photons
that escape from galaxies. With this history in place we then compute
the evolution of the background radiation field due to these same
sources.  After the overlap epoch, ionizing photons will experience
attenuation due to residual overdense pockets of HI gas.  We use the
description of \citet{Miralda2000} to estimate the ionizing
photon mean-free-path [with a reduction of the constant of
proportionality by a factor of 2 following the discussion of
\citet{Oh2005}], and subsequently derive the attenuation of
ionizing photons. We then compute the flux at the Lyman-limit in the
IGM due to sources immediate to each epoch, in addition to redshifted
contributions from earlier epochs. 

\subsection{Star formation and the production of ionizing photons}

In our modeling we assume spectral energy distributions (SED) of
population-II star forming galaxies using the model presented in
\citet{Leitherer1999}. We assume stars of 0.05 solar
metallicity (the effect of metallicity is discussed in \S~\ref{IMF}), and calculate the number ($N_\gamma$) of ionizing photons
above the ionization threshold for hydrogen. This number is calculated
for continuous star-formation, and the low solar metallicity is
reasonable at high redshift. The mass functions each have lower and
upper limits of $0.1$M$_\odot$ and $120$M$_\odot$ respectively. Taking
the functional form $dn/dM\propto M^{-\alpha}$, our fiducial model has
a Salpeter IMF with $\alpha=2.35$ (we also discuss other choices for the
IMF in \S~\ref{IMF}). With these choices we find $N_\gamma=4648$ in
our fiducial model.

\section{Constraints on Models of the Reionization History}
\label{constraints}

In this section we confront our model for reionization with the three
pieces of data described in earlier sections in order to constrain the
possible scenarios.

\subsection{Parameter constraints}

From the above observations we constrain the possible parameterizations of our model for the reionization history. In addition to hypothesised parameters $\Delta_{\rm c}$, $T_{\rm min}$ and $T_{\rm ion}$, our reionization model has two free parameters, $f_\star$ and $f_{\rm esc}$, for combinations of which we compute the reionization history, and calculate the $\chi^2$ of the model as
\begin{eqnarray} 
\nonumber
\chi^2(f_\star,f_{\rm esc})&=&\sum_{i=0}^{N_{\rm \Gamma,obs}}\left(\frac{\log{\Gamma(f_\star,f_{\rm esc})}-\log{\Gamma_{\rm obs}}}{\sigma_{\Gamma}}\right)^2\\
\nonumber
&+&\sum_{i=0}^{N_{\rm \dot{\rho}_\star,obs}}\left(\frac{\log{\dot{\rho}_\star(f_\star,f_{\rm esc})}-\log{\dot{\rho}_{\rm \star,obs}}}{\sigma_{\rho}}\right)^2\\
&+&\left(\frac{\tau_{\rm es}(f_\star,f_{\rm esc})-\tau_{\rm es,obs}}{\sigma_{\rm \tau}}\right)^2.
\end{eqnarray}
Here $\Gamma_{\rm obs}$ and $\dot{\rho}_\star$ are the observed ionization rates and star formation rate densities measured at a number of redshifts, with uncertainty $\sigma_{\Gamma}(r_i)$ and $\sigma_\tau$ (in dex), and $\tau_{\rm es,obs}$ is the observed optical depth to Thomson scattering of CMB photons with uncertainty $\sigma_{\tau}$ (in dex). We note that the error bars on the observational estimates are not symmetric. 

\subsection{Constraints on the escape fraction in the fiducial model}

In the upper left panel of Figure~\ref{fig2}, we present constraints on
the parameters $f_\star$ and $f_{\rm esc}$ assuming our fiducial model
with $T_{\rm min}=10^4$K, and $T_{\rm ion}=10^5$K. As a first
illustration we have included the available constraints on $\tau_{\rm
  es}$ as well as on $\Gamma$, but have neglected $\dot{\rho}_\star$
in order to highlight the importance of its inclusion with the latter
examples. The allowed region in this case is shown by the gray
shading. The ionizing background radiation, and the duration of the
reionization epoch as measured by $\tau_{\rm es}$ constrain the
combination $f_{\rm star}f_{\rm esc}$ \citep{Barkana2001,
  Srbinovsky2008}. This degeneracy can be clearly seen in
Figure~\ref{fig2}.  The dark solid contours that are over plotted in
the upper left panel of Figure~\ref{fig2} show the corresponding
constraints following addition of measurements of $\dot{\rho}_\star$
from $z>4$. The star formation rate breaks this degeneracy, since the
star formation rate density predicted by the model is not directly
dependent\footnote{There is an indirect dependence of star formation
  rate density on $f_{\rm esc}$ because the star formation rate
  depends on the fraction of universe ionized at a particular time.}
on $f_{\rm esc}$. As a result, $f_\star$ and $f_{\rm esc}$ can be
separately constrained yielding values of $0.085\la f_\star\la0.145$
and $0.035\la f_{\rm esc}\la0.060$ for the fiducial model. The best
fit model has the parameter combination ($f_\star,f_{\rm
  esc}$)=(0.110,0.046), with a value of $\chi_{\rm min}^2=1.1$. This
value is total rather than per degree-of-freedom, and is surprisingly
low given a 2 parameter fit and 8 data points which may indicate that
the error bars are overestimated. However we note that the values for 3 additional free parameters have been chosen in this model model ($T_{\rm min}$, $T_{\rm ion}$ and $\Delta_{\rm c}$). Different values
for these are assumed below, leading to larger $\chi^2$ values.

In Figure~\ref{fig1} we show models (solid lines) corresponding to the
best fit, ($f_\star,f_{\rm esc}$)=(0.110,0.046), as well as models
near the edge of the 1-sigma contour, ($f_\star,f_{\rm
  esc}$)=(0.085,0.060) and ($f_\star,f_{\rm esc}$)=(0.145,0.035).  The
locations of these models in the $f_\star$-$f_{\rm esc}$ plane are
shown by dots in Figure~\ref{fig2}. In the upper left and upper right
panels of Figure~\ref{fig1} we show the evolution of the ionization rate and star formation
rate density respectively. In the lower left and lower right panels we show
the evolution of the ionizing photon mean-free-path and mass averaged
neutral fraction that are also predicted by the model. Note that the
quantities other than star formation rate density are degenerate for these
parameter combinations. The data points for mean-free-path are based
on \citet{Storrie-Lombardi1994}.  The observational points for the
mass-fractions are from the damped Ly$\alpha$ measurements of
\citet{Prochaska2005}, and therefore represent lower limits on the
total HI content of the IGM. In both cases the curves show excellent
agreement with these quantities, even though they were not included as
part of the parameter fit. The values of $\tau_{\rm es}$ for these
models are listed in the lower right panel. The universe is predicted
to be 50\% ionized at $z\sim9.5$. 

We note that the observed mean-free-path is found from the number
density of Ly-limit systems and is independent of the Ly$\alpha$
forest absorption derived quantities of ionization rate and volume
averaged neutral fraction, as well as being independent of the HI
mass-density measurements. Our simple model therefore simultaneously
reproduces the evolution of four independent measured quantities, as
well as the optical depth to electron scattering of CMB photons.

Despite this success, the accuracy of estimates for the ionization
rates predicted by the model could be questioned, primarily because of
the sensitivity of the calculation of the mean-free-path to the
probability distribution for $\Delta$. To test the sensitivity of our
conclusion regarding the value of escape fraction to the use of
$\Gamma$ as a constraint, we therefore investigate the constraints
available when only the observations of $\dot{\rho}_\star$ are used in
addition to $\tau_{\rm es}$. The constraints are shown in the upper
right panel of Figure~\ref{fig2}. In this case the best fit model has
the same parameter combination ($f_\star,f_{\rm esc}$)=(0.110,0.046),
with a lower value of $\chi_{\rm min}^2=0.3$.  Thus removing the
$\Gamma$ points loosens the constraints on $f_\star$ and $f_{\rm esc}$
but does not alter the best fit values. We find values of $0.075\la
f_\star\la0.16$ and $0.025\la f_{\rm esc}\la0.10$ in this case.

In Figure~\ref{fig3} we show the marginalised distribution of $f_{\rm
  esc}$. The line styles are consistent with those chosen for the
corresponding contours of $\chi^2$ in Figure~\ref{fig2}. In
constructing the marginalised likelihoods we have assumed prior
probability distributions for $f_\star$ and $f_{\rm esc}$ that are
flat in the logarithm, i.e., $dP_{\rm prior}/d\log{f_\star}\propto1$
and $dP_{\rm prior}/d\log{f_{\rm esc}}\propto1$.

\subsection{Effect of radiative feedback on low mass galaxies}

There is some theoretical uncertainty surrounding the strength of
radiation feedback on the formation of low mass galaxies
\citep{Dijkstra2004b,Mesinger2008}. We therefore repeat our
constraints on $f_\star$ and $f_{\rm esc}$ using a modification of the
fiducial model with no radiative feedback, i.e., $T_{\rm min}=T_{\rm
ion}=10^4$K. The constraints based on this model are shown in the lower left
panel of Figure~\ref{fig2}. The best fit modified model has the
parameter combination ($f_\star,f_{\rm esc}$)=(0.070,0.055), with a
value of $\chi_{\rm min}^2=3.0$ (this model is therefore disfavoured by the data relative to the fiducial case). As reionization progresses, more gas is
allowed to accrete into galaxies in this case than in the fiducial
model. As a result the star formation efficiency required to produce
the observed star formation rate density is reduced. We find values of
$0.05\la f_\star\la0.095$ and $0.04\la f_{\rm esc}\la0.08$ for this
model. Thus the details of the stellar mass accretion rate influence
the inferred star formation efficiency, but only weakly effect the
inferred escape fraction of ionizing photons.

Figure~\ref{fig3} shows the corresponding marginalised distribution
for escape fraction (dashed line). As noted above, the removal of
radiative feedback from the model has little effect on the derived
escape fraction distribution. On the other hand the reionization
history is quite different in this case. The dashed lines in
Figure~\ref{fig1} show the best fit model in this case
($f_\star,f_{\rm esc}$)=(0.07,0.055). The lower value of $f_\star$
means that reionization gets underway later. However despite this
reionization is finished earlier, with an associated rise in the
ionizing background at higher redshifts in this case than in the
fiducial model. This rapid rise occurs because reionization is assumed not
to be self limiting in this case and so is completed by low mass galaxies.

\subsection{Evolving escape fraction}

Our modeling so far has assumed an escape fraction that is constant
with redshift. To test the impact of this assumption on the value of $f_{\rm esc}$ we therefore consider a case where our fiducial model
is modified so that $f_{\rm esc}(z)=f_{\rm esc}(1+z)/7$, and constrain
the value of $f_{\rm esc}$ at $z=6$. The results are shown in the
lower right panel of Figure~\ref{fig2}. In this case we find values
which are very similar to the fiducial case, $0.085\la f_\star\la0.145$
and $0.035\la f_{\rm esc}\la0.06$ for this model. The best fit model
has the parameter combination ($f_\star,f_{\rm esc}$)=(0.11,0.053),
with a value of $\chi_{\rm min}^2=3.1$ (this model is therefore also disfavoured relative to the fiducial case).

\begin{figure}
\includegraphics[width=7.5cm]{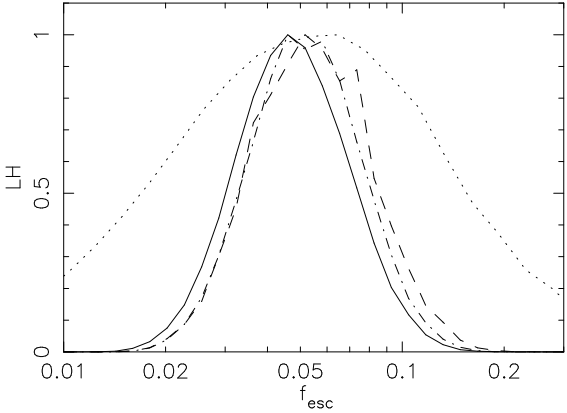} 
\caption{Constraints on the parameter $f_{\rm esc}$ corresponding to marginalisation of the joint distributions in Figure~\ref{fig2} (the line styles are the same for the joint and marginalised distributions in these figures). {\em Solid line:} The fiducial model with $T_{\rm min}=10^4$K and $T_{\rm ion}=10^5$K. Constraints on $\Gamma_{12}$ and $\dot{\rho}_\star$ are used at $z>4$. {\em Dotted line:} Constraints on the fiducial model without inclusion of $\Gamma_{12}$. {\em Dashed Line:} A model with $T_{\rm min}=T_{\rm ion}=10^4$K. Constraints on $\Gamma_{12}$ and $\dot{\rho}_\star$ are used at $z>4$.  {\em Dot-Dashed Line:} A model with $T_{\rm min}=10^4$K and $T_{\rm ion}=10^5$K, and $f_{\rm esc}\propto(1+z)$. Constraints on $\Gamma_{12}$ and $\dot{\rho}_\star$ are used at $z>4$. }
\label{fig3}
\end{figure}

Figure~\ref{fig3} shows the corresponding marginalised distribution
for escape fraction (dot-dashed line). The evolution of escape
fraction is degenerate with evolution in the mass
accretion rate of galaxies, and so as in the previous example has
little effect on the derived escape fraction distribution. The
dot-dashed lines in Figure~\ref{fig1} show the best fit model in this
case ($f_\star,f_{\rm esc}$)=(0.11,0.053). The reionization history is
again quite different from the fiducial model. The larger value of $f_{\rm esc}$
towards high redshift results in an earlier reionization and a higher
$\tau_{\rm es}$.

\subsection{Uncertainty in the choice of IMF}
\label{IMF}

Both the number of ionizing photons produced per star
as well as the star formation rate density inferred from the number of GRBs
observed during the reionization era depend on the IMF assumed. The
former dependence arises because more massive stars are hotter and
therefore emit a larger fraction of their energy at frequencies above
the ionization threshold for hydrogen. Regarding the latter
dependence, we note that the relation between the GRB rate and the
star formation rate density is calibrated at redshifts below $z\sim4$ where
they can each be separately determined \citep{Kistler2009}. Most of
the star formation rate indicators (e.g., H$\alpha$) probe only massive
stars, and hence a correction involving an IMF is required. However,
what is important for the present analysis is the number of ionizing
photons implied by the observed number of high redshift GRBs, i.e., relating one quantity depending just on massive stars to another. With
respect to this quantity the dependencies on IMF mentioned tend to
cancel, because a more top-heavy IMF leads to increases in both the
GRB rate and the number of ionizing photons produced. We note here
that within the context of studies of the star formation history,
{\em top-heavy} can really be interpreted to mean {\em bottom-light}. This is
because one measures the light (or the birth/death rate) from the
massive stars and infers the contribution to stellar mass from lower
masses. In other words the analysis is normalized by the high-mass end
of the IMF.  This situation would be reversed if one were looking at
the integrated stellar mass, which is dominated by low-mass stars.

To assess the systematic error in determination of the escape fraction
introduced by the choice of IMF we consider a range of initial mass
functions (IMF), as summarised in Table ~\ref{tab1}. As in our
fiducial case, we assume stars of 0.05 solar
metallicity, and present the number of ionizing photons above the
ionization threshold for hydrogen. This number is again calculated 
with mass functions that have lower and
upper limits of $0.1$M$_\odot$ and $120$M$_\odot$ respectively.  We
consider two cases with turnover in the mass function below
$0.5$M$_\odot$ (with an index in that mass range of
$\alpha=1.5$). These cases have high mass indexes of $\alpha=2.35$
(labeled Salpeter A) and $\alpha=2.15$ \citep[][labeled BG]{Baldry2003}
respectively. The Salpeter A and BG IMFs were identified by \citet{Hopkins2006} as providing reasonable bounds on the normalisation of
the cosmic star formation history. We also consider the case of a
top-heavy mass function having a single index of $\alpha=1.95$ across
the entire mass range. The values for $N_\gamma$ are listed in
Table~\ref{tab1}, demonstrating that more top-heavy mass functions
produce a larger fraction of their energy as photons above the
ionization threshold. The ratio relative to the value $N_{\rm
  sal}=4648$ for a Salpeter mass function is also presented in the
second last column for ease of comparison.

In the 4th column of Table~\ref{tab1} we show the fraction of the star
formation inferred assuming a Salpeter IMF that would have been
derived in \citet{Kistler2009} if a different IMF had been used
instead. Where a more top-heavy IMF is assumed, the star-formation
rate density implied by the observed GRBs is reduced by this factor relative
to our fiducial model. Since a change in the assumed IMF leads to a
change in the inferred star formation rate density, it also leads to a
proportional change in the derived star formation efficiency. Thus for
each IMF listed the value of $\dot{\rho}_\star/\dot{\rho}_{\rm sal}$
represents the size of the systematic uncertainty in constraints on
$f_\star$.

The change in IMF
also leads to a change in $N_\gamma$, and the escape fraction required
to reproduce the observables of reionization is therefore proportional
to $(\dot{\rho}_\star N_\gamma)^{-1}$. In the final column we list the quantity
\begin{equation}
f_{\rm IMF}\equiv \dot{\rho}_\star/\dot{\rho}_{\rm sal} \times N_\gamma/N_{\rm sal},
\end{equation}
which gives the ratio of ionizing photons to GRBs, relative to the
Salpeter IMF. The range in values of $f_{\rm IMF}$ provides an
estimate of the systematic uncertainty in the derived escape fraction
owing to the choice of IMF. As expected, the net effect on the
reionization considerations caused by choosing a different IMF is
quite modest. It is possible that the IMF is evolving with look back
time. In this case, by extension of the above arguments our estimate
of the escape fraction would sustain only a small systematic
uncertainty.  Importantly, we find that for metal enriched stellar
populations, the choice of IMF contributes a systematic uncertainty on
the derived escape fraction of only $\sim0.1$ dex.

Finally, we note that the assumed metallicity can have an effect on
both the calibration of star formation rate density, and on the value of
$N_\gamma$ computed for a particular IMF. Calibrations of star
formation rate density are performed locally and so refer to solar
metallicity. On the other hand we have assumed a metallicity of 1/20th
solar in computing the reionization history, as may be more
appropriate at high redshift. The Salpeter IMF with solar metallicity
yields $N_\gamma=3328$ ionizing photons per baryon, a factor of 1.4
times smaller than in our fiducial model. This metallicity uncertainty
corresponds to an additional $\sim0.1$ dex uncertainty on $f_{\rm
  esc}$. It may be possible to remove this metallicity uncertainty in
the determination of $f_{\rm esc}$ by noting that many star formation
rate indicators (e.g., H$\alpha$) are proportional to the ionizing
flux, so that the higher number of ionizing photons for low
metallicity stars is offset by a smaller inferred star-formation
rate. We have not pursued this in the current work. We conservatively
estimate that accounting for the uncertainties in IMF and metallicity
leads to $\sim0.2$ dex of systematic uncertainty in the determination
of $f_{\rm esc}$ from our analysis.

\begin{table}
\begin{center}
\caption{\label{tab1} Summary of IMFs used in this work.}
\begin{tabular}{ccccccc}
\hline
 IMF    & $\alpha_{\rm low}$   & $\alpha$ &  $\dot{\rho}_\star/\dot{\rho}_{\rm sal}$   &  $N_\gamma$  & $N_\gamma/N_{\rm sal} $  &  $f_{\rm IMF}$  \\\hline
Salpeter & -  & 2.35 & 1  &  4648 &  1  & 1.00  \\
Salpeter A &  1.5  &  2.35 & 0.77  &  6034 &  1.30  & 1.00  \\
BG03 &        1.5  &  2.15  & 0.50  &  11139 &  2.39  & 1.20  \\
Top-Heavy & -      &  1.95 & 0.37  &  17553 &  2.75  & 1.39  \\\hline
\end{tabular}
\end{center}
\end{table}

\subsection{Reionization by X-rays}

Our analysis has thus far assumed that reionization is entirely due to
high mass stars. However in addition to UV photons from the first
galaxies and quasars, it has been suggested that a background of X-ray
photons at very high redshift may be important
\citep[e.g.,][]{Ricotti2004}. Indeed, several authors have proposed an
initial phase of preheating and partial ionization of the IGM by
X-rays, resulting primarily from black hole accretion
\citep[e.g.,][]{Shull1985,Venkatesan2001,Ricotti2004}. An X-ray
background could also have originated from X-ray binaries and
supernova remnants.  In difference to UV photons, X-rays ionize
hydrogen both directly and through secondary ionizations by
photoelectrons from ionized helium, with the latter dominating so that
many hydrogen atoms can be ionized by a single X-ray photon. As
reionization proceeds, however, the proportion of each X-ray’s energy
that is deposited into the IGM as heat, increases. In particular, for
ionization fractions $\ga10\%$, ionization by secondary electrons
becomes inefficient \citep{Ricotti2004}. As a result, X-ray ionization
is self regulating at the level of $\sim10-20\%$. Modeling the
possible contribution of X-rays to reionization is beyond the scope of
this paper. However the self-regulation implies that neglect of X-rays
would lead to an overestimation of $f_{\rm esc}$ by this factor at
most, and we therefore estimate a systematic component of uncertainty
owing to the unknown X-ray contribution to reionization of
$\sim0.05$ dex.

\subsection{Numerical constraints on the escape fraction}

Our fiducial model constrains the escape fraction to have a value
$\log{f_{\rm esc}}=-1.35\pm0.15$  (68\%). Of the systematic errors discussed
the largest arise due to uncertainty in the IMF and in the
metallicity of the star forming populations responsible for
reionization. We estimate that each of these contribute $\sim0.1$ dex
of systematic uncertainty. Thus in summary of our analysis we find
$\log{f_{\rm esc}}=(-1.35\pm0.15)\pm0.2$.

\section{Summary}
\label{conclusion}

The fraction of ionizing photons that escape their host galaxy and so
are able to ionize hydrogen in the inter-galactic medium (IGM) is a
critical parameter in studies of the reionization era and early galaxy
formation. Indeed as can be seen directly from equation~(\ref{eq1}),
the amount of star formation required to reionize the Universe is
inversely proportional to the escape fraction. Determination of the
expected value for the escape fraction is problematic for individual
galaxies. Observationally, very deep spectra must be obtained to
overcome absorption in the IGM and detect photons that escape the host
galaxy at energies beyond the Ly limit. Theoretical calculation of
escape fraction is difficult owing to the large dynamic range and
complex gas and star formation physics that must be modeled to resolve
the clumpy ISM.

Many studies of the reionization history have compared theoretical
models to the electron scattering optical depth of CMB photons, and
the ionization rate from the Ly$\alpha$ forest at the end of the
overlap era. These studies are subject to degeneracies between the
escape fraction of ionizing photons, and the efficiency with which high redshift
galaxies turn mass into stars
\citep[e.g.,][]{Wyithe2003,Haiman2003,Cen2003,Choudhury2005,Shull2008}. While upper limits have previously been obtained for surveys of high redshift star forming galaxies \citep{Bouwens2008}, the
recent determination of the star formation rate density during the
reionization era from the most distant Gamma Ray Bursts discovered
provides the first method for directly measuring the star formation
rate density during the reionization era. With respect to calculating the
escape fraction, this method has the advantages over estimates of star
formation rate from the galaxy luminosity function that it directly
observes the total star formation rate (i.e., no correction for a
galaxy flux limit), and is well calibrated at lower redshift
\citep{Kistler2009}.

In this paper we have used this first determination of the star
formation rate density at $z>6$ to break the degeneracy between star formation
efficiency and escape fraction.  Our analysis employed a semi-analytic
model for reionization that we confronted with three complementary
sets of observations, $i)$ the electron scattering optical depth of
CMB photons, $ii)$ the ionization rate from the Ly$\alpha$ forest at
the end of the overlap era and $iii)$ The star formation rate density out to
$z=8.5$. By constraining the model parameters with these observations
we show that the escape fraction of ionizing photons from high
redshift galaxies is $f_{\rm esc}\sim5\%$. For our fiducial model we find $\log{f_{\rm esc}}=-1.35\pm0.15$

Importantly, this value is stable against a range of systematic
uncertainties in the modeling, including the implementation of
radiative feedback and the choice of IMF.  In agreement with previous
studies we find that the mass assembly history of galaxies was
important for the reionization history, and impacts on the inferred value of
star formation efficiency. However, the value of escape fraction
inferred is not sensitive to the details of the mass assembly. We
also investigated an evolving escape fraction, finding that this did
not strongly influence the value of escape fraction inferred for
galaxies at $z=6$. Our modelling does make the assumption that UV
photons produced by stars dominate the reionization process. A
possible contribution of X-rays to reionization \citep{Ricotti2004}
complicates this interpretation, and implies that our escape fraction
could be overestimated by a factor up to $1.1-1.2$. Both the number of ionizing
photons and the number of GRBs per unit star formation are sensitive
to the IMF. The dependencies on IMF tend to cancel in our analysis,
however we considered a range of possibilities. We estimate an
uncertainty of $\sim0.1$ dex in $f_{\rm esc}$ owing to the unknown shape
of the IMF, and another $\sim0.1$ dex owing to the unknown metallicity of
the galaxies responsible for reionization. Including these
uncertainties we express our estimate of escape fraction as $\log{f_{\rm esc}}=(-1.35\pm0.15)\pm0.2$. Our analysis assumed that
the escape fraction is independent of galaxy mass. Previous work has
suggested that lower mass galaxies have smaller escape fractions
\citep{Gnedin2007a}. In this scenario our escape fraction should be
considered as a stellar mass weighted average over star forming galaxies.

Our simple semi-analytic model is able to provide a statistically
acceptable fit to the available observations under a range of
assumptions. This implies that models of the reionization history are
under constrained [although more variables could potentially be used
as constraints \citet{Choudhury2005,Choudhury2006}]. In the near
future 21cm observations of neutral IGM during the reionization era
will provide a powerful new probe. \citet{Lidz2008} have shown that
the first generation of low frequency telescopes will be able to
measure the neutral fraction accurately during epochs when the
universe is $\sim50\%$ ionized. The example reionization histories
presented in this and other papers illustrate that this will provide
important discriminates among possible reionization histories. With
respect to the escape fraction, such measurements in combination with
the evolution in star formation rate density will allow for the evolution in
addition to the value to be measured.


\section*{Acknowledgements}
This work was supported in part by the Australian Research Council
(JSBW and AMH).  MDK and JFB were supported by NSF CAREER Grant PHY-0547102 (to JFB) and HY by DOE Grant DE-FG02-91ER40626.

\bibliographystyle{mn2e}
\bibliography{text}
\end{document}